\documentclass[twocolumn,aps,prl,groupedaddress]{revtex4-1}
\usepackage{graphicx}
\usepackage{color}
\usepackage{amsmath}
\usepackage{enumitem}
\usepackage{amssymb}
\usepackage{hyperref}

\usepackage{bm}
\usepackage{subfigure}
\usepackage[normalem]{ulem}

\newcommand{\be}{\begin{equation}}
\newcommand{\ee}{\end{equation}}

\newcommand{\bea}{\begin{eqnarray}}
\newcommand{\eea}{\end{eqnarray}}
\newcommand{\HH}{{\cal H}}

\newcommand{\la}{\left\langle}
\newcommand{\ra}{\right\rangle}

\renewcommand{\vec}[1]{{\boldsymbol #1}}

\newcommand{\addLL}[1]{\textcolor{blue}{#1}}

\begin{document}
\title{Electric-Dipole Spin Resonances}

\author{E. I.  Rashba${}^1$, V. I. Sheka${}^2$}
\affiliation{$^1$L.D. Landau Institute for Theoretical Physics of the USSR Academy of Sciences, 
117940 GSP-I Moscow, ul Kosygina 2, USSR\\ $^2$ Institute of Semiconductors of the Academy of Sciences of the Ukranian SSR, 252028 Kiev, prospekt Nauki 45, USSR}
%
%\date{\today}
%
\begin{abstract}
Resonance phenomena in solids generally fall into two distinct classes, electric and magnetic, driven, respectively, by the $E$ and $H$ components of the electromagnetic wave incident on the solid. The canonical examples of the two types of resonances are the electron cyclotron resonance (CR) and the electron paramagnetic resonance (EPR), originating from the electron orbital and spin degrees of freedom, respectively. The behavior becomes considerably more interesting (and more complicated) in the presence of the spin-orbital interaction. In this case, a more general type of resonance may occur, which is driven by the electric excitation mechanism and involves 
the spin degrees of freedom. Such electric-dipole spin resonance (EDSR) may occur at the spin excitation frequency or at a combination of the orbital and spin frequencies, spanning a wide bandwidth. The EDSR phenomenon, first predicted by Rashba (1960), has been probed experimentally  in 3D solids with different crystal symmetries, as well as in low-dimensional systems (heterojunctions, inversion layers, dislocations and impurity states). Due to its electric dipole origin, the EDSR features a relatively high intensity, which may exceed by orders of magnitude the EPR intensity. This review summarizes the work on EDSR prior to 1991, laying out the theoretical framework and discussing different experimental systems in which the EDSR-related physics can be realized and explored.
\end{abstract}
%
%% insert suggested PACS numbers in braces on next line
%%\pacs{}

\maketitle

%The published version of this article is available here: %at these links: 
%\href{http://www.mit.edu/~levitov/rashba_papers/1991RashbaSheka.pdf}{\addLL{PDF}}
%(human-friendly), \href{http://www.mit.edu/~levitov/rashba_papers/1991RashbaSheka.pdf}{\addLL{PDF}}
%(robot-friendly).

%\tableofcontents

\medskip

{\it  Contents:}
 
1.	Introduction %....................................................................... 133

2.	Basic formalism of the theory %...................................................... 137

3.	COR theory in the Zeeman limit %..................................................... 140

4.	Angular %indicatrices 
distributions and selection rules %..........................................  144

5.~Three-dimensional spectrum with linear terms in the dispersion law %................. 150

6.	Inversion asymmetry mechanism for the n-type InSb band %............................. 157

7.	EDSR and EPR interference %.........................................................  161

8.	COR for semiconductors with inversion centre %....................................... 164

9.	COR for narrow-gap and zero-gap semiconductors %..................................... 167

10.	COR on shallow local centres %........................................................ 172

11.~Two-dimensional systems: heterojunctions and MOS structures %......................... 178

12.	One-dimensional systems: dislocations %.............................................. 181

13.	Shape of the EDSR band %............................................................. 184

14.	EDSR induced by lattice imperfections %.............................................. 188

15.	Conclusion %......................................................................... 191

Addendum A. Transformation of the reference system and of the Hamiltonian %............... 193

Addendum B. Kane model %................................................................   195

List of abbreviations %................................................................... 202

References %.............................................................................. 202

\section{Introduction}

Resonance phenomena, which offer a powerful tool of studying intricate properties of condensed matter, have for a long time been divided into electric and magnetic resonances. Electron cyclotron resonance (CR), for instance, belongs to the class of electric resonances, whereas electron paramagnetic resonance (EPR) belongs to the class of magnetic resonance phenomena. Electric resonances are excited by the electric vector of an electromagnetic wave, and electrons experience a change in their orbital state but not in their spin state.

Naturally, the classification of the motion in terms of the coordinate and spin degrees of freedom, which in fact provides the grounds for such a simplified description, is possible only in the absence of spin orbit (SO) interaction. The subject of this paper is an electron resonance of a more general type, excited due to SO interaction. The characteristic features of this resonance, called combined resonance (COR), are: (i) the electric mechanism of its excitation, and (ii) change of the spin quantum state, when the quantum numbers corresponding to the orbital motion either remain unchanged or are changing. In the former case the resonance occurs at the spin frequency of an electron and is called electric-dipole spin resonance (EDSR), or electric-dipole-excited electron-spin resonance (EDE-ESR). In the latter case it occurs at combinational frequencies, that is, linear combinations of orbital and spin frequencies. We shall call this the electric-dipole combinational frequency resonance. If the mechanism of its excitation is not specified, or if the emphasis is on the frequency of the resonance rather than on the mechanism of its excitation, we shall use the terms spin resonance (SR) or combinational frequency resonance (CFR).

At present COR, first predicted by Rashba (1960, 1961), is being experimentally discovered and studied for various crystals with different types of symmetry. It has been observed in 3D systems, i.e. in bulk, in 2D systems (on heterojunctions and inversion layers), in ID systems (on dislocations) and in 0D systems (on impurity centres). Now COR is regularly used to investigate the band structure of semiconductors. The extensive use of the method is accounted for by: (i) the relatively high intensity of COR (which may exceed the EPR intensity by several orders of magnitude), (ii) the presence, as a rule, of several COR bands in the spectrum, and (iii) the fairly specific angular dependence of their intensity.

Now it is necessary to clarify, first, what the source is of the high COR intensity, and, second, why for more than 15 years after the discovery of EPR by Zavoisky (1945) and its extensive experimental investigation, COR was not observed. It is convenient to start this discussion by taking band electrons as an example.

In the absence of SO interaction, an electron put in a constant homogeneous magnetic field $H$, performs two independent motions associated with orbital and spin degrees of freedom. The first motion is cyclotron rotation with a frequency
\[(1.1)\qquad 
\omega_c=eH/m^*c
%(1.1)
\]
where $m^*$ is the effective mass, and $e$ and $c$ are universal constants. A characteristic spatial scale corresponding to this motion is the magnetic length
\[(1.2)\qquad
r_H=(c\hbar/eH)^{1/2}
%(1.2)
\]
Accordingly, the minimal electric-dipole moment corresponding to a transition between neighbouring quantum states under the influence of an a.c. electric field is $p_c\approx er_H$. The frequency $\omega_s$ of spin transitions is determined by the equation
$\omega_s=|g|\mu_BH$, where $g$ is the $g$-factor of an electron in a crystal and $\mu_B=e\hbar/2m_0c$ is the Bohr magneton, and $m_0$ is the mass of the electron in a vacuum. Putting $g \sim 2$, we can estimate the magnetic-dipole moment of the transition displayed in EPR as $\mu_s\sim e\lambda$. Here $\lambda$ is the Compton electron wavelength: $\lambda\approx 4\times 10^{-11}$cm.
Therefore, if the values of $\tilde E$ and $\tilde H$ (the amplitudes of the electric and magnetic fields, respectively) are comparable, the ratio of the CR and EPR intensities is of the order of $I_{\rm CR}/I_{\rm EPR}\sim (r_H/\lambda)^2$. For typical values of $H$, $r_H \sim 10^{-5}-10^{-6}$cm. This allows us to estimate the value of the ratio of the intensities; typically, it can be as large as $I_{\rm CR}/I_{\rm EPR}\sim 10^{10}$.

So, CR is many orders of magnitude stronger than EPR. This property is inherent in all electric resonances. For instance, for paraelectric resonance, $r_H$ must be replaced by the characteristic atomic quantity, namely, the Bohr radius $r_B =\hbar^2/m_0e^2 = 0.5\times 10^{-8}$cm, and therefore the ratio of the intensities is $I_{\rm CR}/I_{\rm EPR}\sim (r_h/\lambda)^2 \sim 10^4$. Since electric resonances are much stronger than magnetic resonances, one can expect that even weak SO interaction leading to the coupling of orbital and spin motions will cause intensive electric excitation of SR. Besides, for band electrons the coupling of orbital and spin motions makes it possible for the combinational frequencies $\omega = n\omega_c \pm \omega_s$ (where $n$ is an integer) to appear in the spectrum. The intensity of the transition at these frequencies, i.e., the intensity of the electric-dipole CFR, is generally of the same order of magnitude as the EDSR intensity. Jointly they form the COR spectrum. 

%\addLL{
We note parenthetically that very often only the electric-dipole CFR bands are ascribed to COR. However, we shall use the term COR in the sense defined above, in conformity with the original work (Rashba 1960) and with subsequent reviews (Rashba 1964a, 1979). Thus by COR we understand the entire family of electric-dipole spin resonances.

Usually it is convenient to observe the COR spectrum in cyclotron-resonance inactive (CRI) polarizations, since there is no strong CR background.
Now let us clarify why COR may be absent or, more exactly, very weak. Let us start with a free electron in a vacuum. The Thomas SO interaction energy is
\[(1.3)\qquad
\HH_{\rm so}=(\mu_B/2m_0c){\boldsymbol\sigma}\cdot (\vec E\times\vec p)
%(1.3)
\]
%\footnotetext{\addLL{Very often only the electric-dipole CFR bands are ascribed to COR. However, we shall use the term COR in the sense defined above, in conformity with the original work (Rashba 1960) and with subsequent reviews (Rashba 1964a, 1979). Thus by COR we understand the entire family of electric-dipole spin resonances.}}
Here ${\boldsymbol\sigma}=(\sigma_x,\sigma_y,\sigma_z)$ are Pauli matrices, $\vec E$ is the electric field and $\vec p$ is the momentum operator. If $\vec E$ is regarded as an a.c. electric field $\tilde{\vec E}$ with the frequency $\omega_s$ and $\vec v = \vec p/m_0$ is taken as the velocity of the electron, then $\HH_{\rm so}=\frac12\mu_B {\boldsymbol\sigma}\cdot (\vec E\times\vec v/c)$. Comparing this expression with the Zeeman energy $\mu_BH$, and assuming $\tilde E =\tilde  H$, we see that $I_{\rm EDSR}/I_{\rm EPR}\sim (v/c)^2$. In the nonrelativistic limit, $(v/c)^2 \ll 1$ and EDSR is much weaker than EPR. This result is absolutely clear since up to the `Thomas 1/2' the energy $\HH_{\rm so}$ coincides with the Zeeman energy in the effective magnetic field $H_{\rm eff} = (v/c)E$ which acts on the electron in the reference system where the electron rests. Therefore the SO interaction is required to be sufficiently strong. In crystals, SO interaction is strong due to the fact that the field $\vec E$ in (1.3) is a static electric field of the crystal lattice which is particularly strong near nuclei and the operator $\vec p$ acts not on the smooth functions of the effective mass approximation 
%(EMA) 
but on the full Bloch functions:
\[(1.4)\qquad
\psi_{nk}(\vec r)=u_{nk}(\vec r)\exp(i\vec k\cdot \vec r)
%(1.4)
\]
the periodic factor $u_{nk}(\vec r)$ rapidly varies near nuclei. As a result, the SO interaction becomes stronger with increasing charges of the nuclei of the atoms constituting the crystal. In typical semiconductors the SO splitting of the valence band is $\Delta\sim 0.1$-$1$eV and it may compete with the forbidden gap width $E_G$. Thus the $g$-factor and other parameters of the electron are strongly renormalized as compared to the parameters of an electron in a vacuum. For example, the $g$-factor may change substantially and may have an anomalous sign (Yafet 1963). As a result, the spin of the electron somehow transforms into its ‘quasispin'. Then due to the difference between $\mu^*=g\mu_B/2$ and $\mu_B$ the EPR intensity varies: at larger values of $|g|$ it may be much higher than for an electron in vacuum. But the COR intensity is determined not by the renormalized value of the $g$-factor but by specific terms in the EMA Hamiltonian which simultaneously involve the Pauli matrices (quasispin) and the quasimomentum operator $\vec k$ (orbit). The structure of these terms and, consequently, the COR intensity is determined by the symmetry of the crystal. This problem is discussed in section 2 and subsequent sections. On the whole, the higher the symmetry of the group Gk of the wave vector in the point of $\vec k$-space corresponding to the band extremum, and the larger the $E_G$, the lower the COR intensity. Naturally, the intensity decreases with a decreasing charge of the nucleus.

Despite the aforementioned restrictions the COR intensity for band carriers in many crystals is so high that it is impossible to observe EPR against the background of the COR intensity. However it significantly decreases in cases where electrons become bound in donor states. It follows from the Kramers theorem (see section 10) that in this case the intensity involves the factor $(\hbar\omega_s/{\cal E}_i)^2$, where ${\cal E}_i$ is the ionization potential of a donor. This factor can be very small if the field $H$ is weak.

Sticking to the subject of this volume, we shall consider only COR of band carriers in the Landau levels and also electrons bound to shallow impurity centres where ${\cal E}_i$ is comparable with $\omega_c$ and $\omega_s$. But it should be noted that EDSR is possible also for low-symmetry deep centres. For them, the EDSR intensity is determined not by the band spectrum of the semiconductor but by the structure of the electron shell of an impurity ion and by the local symmetry of the crystal field. On the whole, it is noticeably lower than for band electrons and for large-radius centres. EDSR for small-radius centres was predicted by Bloembcrgen (1961) and experimentally observed by Ludwig and Ham (1962). The survey by Roitsin (1971) and two monographs by Mims (1976) and by Glinchuk et al. (1981) are devoted to electric effects in the radiospectroscopy of deep centres.

The foregoing arguments shed some light upon why EDSR was not observed and identified experimentally until the conditions of its high intensity had been found theoretically.

The COR mechanism for free carriers we have discussed is totally due to the SO interaction entering in the Hamiltonian for a free carrier in a perfect crystal. According to this approach the presence of impurities or defects, which cause binding of carriers in shallow levels, diminishes the COR intensity. The theory based on this concept was developed in the early sixties and preceded the experiment: its results were summarized in a survey by Rashba (1964a). The diverse experimental data obtained since then are in agreement with the theory, and have permitted a number of new parameters of the energy spectrum of carriers to be found. Later theoretical works were aimed at describing the experimental results quantitatively within the framework of the original concept.

However, there is one other line of thinking in COR physics. Originating from experimental results rather than from theoretical ideas, its essence is the existence of specific COR mechanisms, caused by defects or impurities. As a result, in materials with a high concentration of imperfections, COR may prove to be considerably stronger than in high-quality samples. The paper by Bell (1962) on EDSR in strongly doped n-type InSb was the first to point to the existence of such mechanisms, and the problem was first recognized and formulated by Mel'nikov and Rashba (1971). For the moment, the problem remains somewhat obscure. That is why there is no doubt that future work on COR theory must be concentrated on this problem. One can expect the problem to attract the attention of experimentalists since it opens up new possibilities for studying disordered systems.

In terms of macroscopic electrodynamics, COR belongs to magnetoelectric phenomena, first reported by Curie (1894) and reviewed by O'Dell (1970) for magnetic materials. From the viewpoint of the microscopic mechanism the most significant feature of COR is the strong coupling of electron spins to the a.c. electric field in a broad class of crystals. By now different manifestations of this coupling have been found. This coupling, in particular, is responsible for spin-flip Raman scattering, discussed in chapter 5 by Hafele.

\section{Basic formalism of the theory}

The COR theory is based on the theory of the band spectrum of an electron in crystals and on the effective mass approximation. These concepts have been thoroughly developed and they were reviewed, for instance, in the paper by Blount (1962) and in the book by Bir and Pikus (1972), which we recommend to the reader.

In practical COR calculations for specific semiconductors one should proceed from the band structure of the semiconductor determined by: (i) symmetry properties and, (ii) by numerical values of the energy spectrum parameters. For example, for semiconductors with a narrow forbidden gap of the InSb-type it is often convenient to make use of a multiband Kane model (1957). However, (i) to elucidate the principal mechanisms of the COR phenomenon and, (ii) to do it from the same point of view conformably to different systems, it will be more convenient to use a two-branch (i.e. one-band) model wherever possible. By this term we understand two branches of the energy spectrum differing only in the spin state of an electron (or a hole). These branches of the spectrum in crystals with the inversion centre merge into one band in the entire $\vec k$-space (Elliott 1954) and in crystals without the inversion centre they stick together in a high-symmetry point and split in its vicinity. Numerical parameters of the two-branch model can be expressed via parameters of a more general model (Addendum B).

In the framework of the two-branch model, the most general approach to describe COR in a semiconductor, subjected to external fields (electric and magnetic), is as follows. The Hamiltonian $H$ for an electron and the operator $\hat{\vec r}$ of its coordinate can be derived by means of the method of invariants (Bir and Pikus 1972) which relies only on general symmetry requirements:
\[(2.1)\qquad
\HH=\HH_0+e\varphi(\hat{\vec r})+\HH_{\rm so}, \quad
%(2.1)
\]
\[(2.2)\qquad
\HH_{\rm so}=\sum_j\sigma_jf_j(\hat{\vec k}), \quad
%(2.2)
\]
\[(2.3)\qquad
\hat{\vec r}=i\nabla_{\vec k}+\sum_j \sigma_j X_j(\hat{\vec k})
%(2.3)
\]
where $\sigma_j$ are Pauli matrices and $\hat{\vec k}$ is the operator of the magnetic field quasimomentum
\[(2.4)\qquad
\hat{\vec k}=-i\nabla_{\vec r}-(e/c\hbar)\vec A(\vec r),
%(2.4)
\]
$\vec A(\vec r)$ and $\varphi (\vec r)$ are the vector and scalar potentials, respectively. The functions $f_i(\hat{\vec k})$ and $X_i(\hat{\vec k})$ are polynomials over powers of $\hat{k_j}$, the Cartesian coordinates of the $\hat{\vec k}$ vector. These polynomials are such that $\HH$ and $\hat{\vec r}$ possess the necessary transformation properties with respect to $G_{\vec  k}$, the little group of $\vec k$, the wave vector near which an expansion in powers of $k_j$ is performed. Namely, $\HH$ must behave as a scalar quantity and $\hat{\vec r}$ as a vector quantity with respect to spatial transformations. Both $\HH$ and $\hat{\vec r}$ must be real operators, i.e., must retain their sign upon time reversal $t\to -t$. The functions $f_i$ and $X_i$ include both symmetrized and antisymmetrized combinations of $\hat{k_j}$. %In virtue of 
Owing to the commutation condition
\[(2.5)\qquad
[\hat k_j,\hat k_{j'}]=i\frac{e}{c\hbar}H_{j''}
%(2.5)
\]
where $j$, $j'$ and $j''$ constitute cyclic permutations (e.g., if $j= 2$ or $j = y$, then $j' = 3$, $j'' = 1$, or $j = z$, $j" = x$), the antisymmetrized terms can be expressed via $H_j$. The potential $\varphi(r)$ can be describe, for instance, the potential created by impurities. It will be assumed that this potential is smooth. %In the higher \addQ{EMA} 
Within the effective mass approximation, at a higher order, alongside $\varphi(r)$ there emerges a gradient $\vec E= -\nabla_{\vec r}\varphi(r)$ in $\HH$. The corresponding term in $\HH$ is analogous to the SO interaction (1.3) for a Dirac electron (eigenfunctions of the operator $\HH$ are two-component spinors).

If the a.c. electric field $\tilde{\vec E}$ exciting resonance transitions is described by the vector-potential $\tilde{\vec A}$, the interaction Hamiltonian is
\[(2.6)
\qquad
\HH_{e}=-(e/c)\hat{\vec v}\cdot\tilde{\vec A}
\]
where the velocity operator is determined by a commutator
\[
(2.7)\qquad
\vec v=\frac{i}{\hbar}[\HH, \hat{\vec r}].
\]
From eqs. (2.1) -(2.3) and (2.7) it follows that
\[(2.8)\qquad
\vec v=\hbar^{-1}\nabla_{\vec k}\HH+{\boldsymbol\Omega}(\hat{\vec k})
\]
where the vector  ${\boldsymbol\Omega}(\hat{\vec k})$ is the polynomial over $\hat{\vec k}$. Note that the $X_j(\vec k)$ operators become important only when the terms of the order $k^4$ or higher are taken into account
in $f_j(\vec k)$. 

A complicated structure of the $\vec r$ operators in eq. (2.3) results from projecting the multiband Hamiltonian of the $\vec k\vec p$ approximation (Luttinger and Kohn 1955) onto the conduction band (or valence band). Similar terms also exist in the Dirac problem. The Dirac Hamiltonian may be treated as a multiband Hamiltonian, simultaneously incorporating dynamics of differently charged particles (electrons and positrons, or, in terms of the solid state theory, electrons and holes). From this point of view, interband matrix elements must correlate as $c\vec p = c\hbar \vec k$ $\to$ $P\vec k$ and the `forbidden energy gap' as $2m_0c^2$ $\to$ $E_g$. In the $1/c^2$ approximation, $\hat{\vec r}$ becomes
\[(2.9)\qquad
\hat{\vec r}=\vec r+\frac{\hbar}{4m_0c^2}({\boldsymbol \sigma}\times\hat{\vec p})
\quad
\to
\quad
i\nabla_{\vec k}+\frac{P^2}{E_g^2}({\boldsymbol \sigma}\times\hat{\vec k})
.
\]
For semiconductors the coefficient entering in the SO term is much larger than the %appropriate 
corresponding coefficient in a vacuum, as has been pointed out in section 1. 

The interaction of an electron with an electromagnetic wave can be described not only by the vector but also by the scalar potential $\varphi(\vec r) = -e\vec E\vec r$. Then the COR intensity is expressed via matrix elements of the $\hat{\vec r}$ operator. Due to the relation
\[(2.10)\qquad
\la f|\hat{\vec v}|i\ra=i\omega_{fi}\la f|\hat{\vec r}|i\ra
%<f|B|i> = icufi<f|r|i>	
%(2.10)
\]
($\omega_{fi}$ is the transition frequency), straightforwardly following from (2.7), the results obtained by either method coincide. It is worth %stressing 
noting that the matrix elements $\la f|\hat{\vec r}|i\ra$ depend not only on $\vec X_j$ but also on $\HH_{\rm so}$. This is because the wave functions of the $i$ and $f$ states are also $\HH_{\rm so}$-dependent, this dependence being quite relevant (Rashba and Sheka 1961a, c).
To compare the EDSR and EPR intensities, it is necessary to calculate matrix elements of the interaction responsible for EPR. They are determined by the magnetic component of the electromagnetic field. The corresponding perturbation operator equals
\[(2.11)\qquad
\tilde{\HH}_{\rm m}=\tilde{\vec H}\cdot\nabla_{\vec H}\HH=(\nabla\times\vec A)\cdot\nabla_{\vec H}\HH
%(2.11)
\]
Differentiation in (2.11) should be performed only with respect to $\vec H$ entering explicitly in $\HH$ but not with respect to $\vec H$ entering through the vector-potential $\vec A(\vec H)$, since the %appropriate 
corresponding terms are already taken into account in (2.6).

Experiments typically use two types of mutual orientation of the unit vector $\vec e$ of the electric field of an electromagnetic wave, of its wave vector $\vec q$ and of the constant magnetic field $\vec H$: the Faraday geometry ($\vec q\parallel \vec H$, $\vec e\perp \vec H$) with two circular polarizations of $\vec e$ (transverse resonance), and the longitudinal Voigt geometry ($\vec q\perp \vec H$, $\vec e\parallel \vec H$) (longitudinal resonance). This choice of polarizations is also handy for constructing the theory. Therefore, apart from the original reference system A, associated with the crystallographic axes, it is useful to introduce another reference system A', associated with the magnetic field in such a manner that $ Z\parallel\vec H$ (Addendum A). In the A system the Cartesian basis is employed and the vectors $\vec r$, $\vec k$ and $\vec v$ are denoted by lowercase letters and their coordinates are numbered by Latin indices ($i, j= 1, 2, 3$). In the A' system the vectors are denoted by capital letters. Their components are chosen in the circular basis
\[(2.12)\qquad
\vec V=(V_-,V_Z,V_+)=(V_{\bar 1},V_0,V_1), 
%y=(V_, Vz, K+) = (Fb У0, K,),	(2.12)
\]
and similarly for $\vec R$ and $\vec K$. In (2.12)
\[(2.13)\qquad
V_{\pm }=(V_X\pm iV_Y)/\sqrt{2}
%F± = ( F* ± i Vy)ijl	(2.13)
\]
In the circular basis the coordinates are numbered by Greek indices $\alpha,\beta = \bar 1, 0, 1$ or $-1,0, 1$.

The direction of the $Z$-axis will be chosen from the condition $eH_Z > 0$ with the sign of the charge $e$ taken into account. According to the conditions (2.5) $K_Z$ is a c-number and the other components obey the commutation rules
\[(2.14)\qquad
[\hat K_-,\hat K_+ ] = eH/c\hbar \equiv  k_H^2
,\quad
 k_H=r_H^{-1}.
%(2.14)
\]
If we %single 
factor out the dimensional factor $k_H$ from $\hat K$, the result can be represented by %step-up and step-down 
the raising and lowering ladder operators $a^+$ and $a$ as
\[(2.15) \quad
\hat{\vec K}=k_H\vec a,\quad
\vec a=(a,\xi,a^+),\quad
[aa^+-a^+a]= 1,\quad
%\xi=k_H^{-1}K_z
%K = kna, a = (a,Ç,a% aa+-a+a= I, i = k#*Kz. 
%(2.15) 
\]
$\xi=k_H^{-1}K_z$. In the circular basis the commutators of $\hat{\vec K}$ and $\vec R$ are written as
\[(2.16)\qquad
[\hat K_\alpha, R_\beta]=-i\delta_{\alpha\beta}
%i<V	
%(2.16)
\]
So far we have dealt with purely technical aspects of the problem relevant to the formalism of calculations. In conclusion to this section we shall make an attempt to discuss the problem in physical terms. This will enable us to understand qualitatively certain COR mechanisms. Of course, such considerations are no substitute for a consistent analysis of the Hamiltonian (2.2) or for a more sophisticated Hamiltonian describing the multiband model.

There is a qualitative distinction between band structures of crystals with an inversion centre and crystals without one. In this section it has already been noted that in crystals without an inversion centre the spectrum is degenerate only at certain points of the $\vec k$-space, but in the vicinity of these points the degeneracy is lifted and the spectrum splits into two branches corresponding to different spin states of the electron. This splitting is due to the Hamiltonian (2.2) where $f_j$ are linear or cubic in $k$ (cf. sections 5 and 6). Since such terms in $\HH_{\rm so}$ are inherent in crystals without an inversion centre, the COR excitation mechanism caused by them is termed the inversion asymmetry mechanism. As a rule, it is fairly efficient.

In crystals with an inversion centre, $f_j\propto H$. This is indispensable for ensuring twofold degeneracy of bands for all $\vec k$. But the presence of the $H$ factor diminishes $\HH_{\rm so}$ and it may have an observable value only if the forbidden energy gap $E_g$ entering in the denominator of $f_j$ is narrow. Under these conditions the region, where the dependence of $f_j$ on $\vec k$ is quadratic, will be very narrow; this is why the COR mechanism associated with a small value of $E_g$ is often termed the nonparabolicity mechanism (see section 8). Sometimes under these conditions a major role is played by the large value of the $X_j(\vec k)$ functions, a possibility made clear from (2.9) (see also section 9).

Above we have covered the two mechanisms which can be most clearly specified. But in realistic situations, especially when one is dealing with degenerate valence bands, to distinguish and interpret individually the contributions of different perturbations (in particular, of those responsible for warping) is practically impossible.

\section{COR theory in the Zeeman limit}

An exact analytical solution of the problem for an electron in a homogeneous magnetic field Я can be derived only for a few specific cases even for the two-branch model. Yet, the most interesting situation occurs when the Zeeman splitting dominates over SO splitting (Zeeman limit). It can be studied in the general form at $\varphi(\vec r) = 0$. In this case an expansion is performed in the parameter
\[(3.1)\qquad
\gamma(\bar k)\approx \la \HH_{\rm so}^2\ra^{1/2}/\hbar\omega_{\rm min}\ll 1
%y(C) ss C^ty1!2/hcomi„ « I,	
%(3.1)
\]
where $\omega_{\rm min} = {\rm min}\{\omega_c, \omega_s, n\omega_c-\omega_s\}$, $n$ is an integer. Here $\bar k$ is a characteristic value of the quasimomentum; for instance, for a band electron it is determined by formula (3.4). The criterion (3.1) means that the mean energy of SO interaction is small compared to the spacing between magnetic quantization levels. Depending on the power $l$ of the quasimomentum $\vec k$ entering in $\HH_{\rm so}$, the criterion (3.1) is fulfilled in strong $(l= 1)$ or weak $(l\ge 3)$ fields.

\medskip

\centerline{[followed by about $30$ more pages]}

\medskip

The published version of this article that is recommended to the reader is available here: %at these links: 
\href{http://www.mit.edu/~levitov/rashba_papers/1991RashbaSheka.pdf}{\addLL{PDF}}
(human-friendly), \href{http://www.mit.edu/~levitov/rashba_papers/rashba_ocr.pdf}{\addLL{PDF}}
(robot-friendly).  Some of the %historic 
articles cited in the bibliography, which are not easily available online, can be found \href{http://www.mit.edu/~levitov/rashba_papers/}{\addLL{here}} (active link).

 \begin{figure}[hbt!]
\includegraphics[width=0.6\columnwidth]{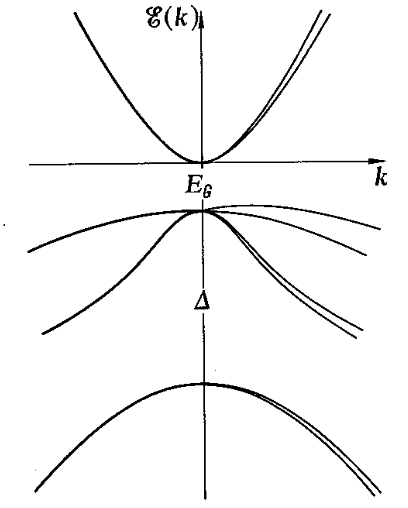} 
\caption{Arrangement of bands in direct-gap cubic semiconductors, described by the Kane model (1957). $E_g$ is the width of the forbidden band, and A is the SO splitting. On the top is the conduction band; in the middle the valence band, consisting of the light hole and heavy hole bands; at the bottom is the split-off band. The figure describes %applies to 
the ${\rm A_{III} B_V}$-type %of 
semiconductors. The weak splitting of the bands (emphasizing their twofold degeneracy) is due to the absence of an inversion centre. On the left-hand side, the splitting is neglected; on the right-hand side, the splitting is shown but its magnitude is
exaggerated.
} 
\label{fig1}
%\vspace{-4mm}
\end{figure}

 \begin{figure}[hbt!]
\includegraphics[width=0.95\columnwidth]{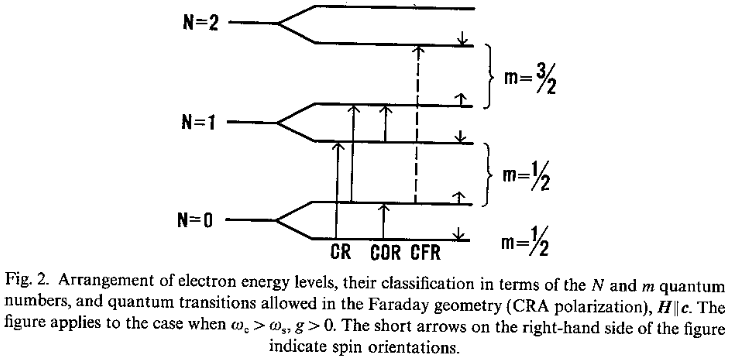} 
\caption{ 
} 
\label{fig2}
%\vspace{-4mm}
\end{figure}

 \begin{figure}[hbt!]
\includegraphics[width=0.95\columnwidth]{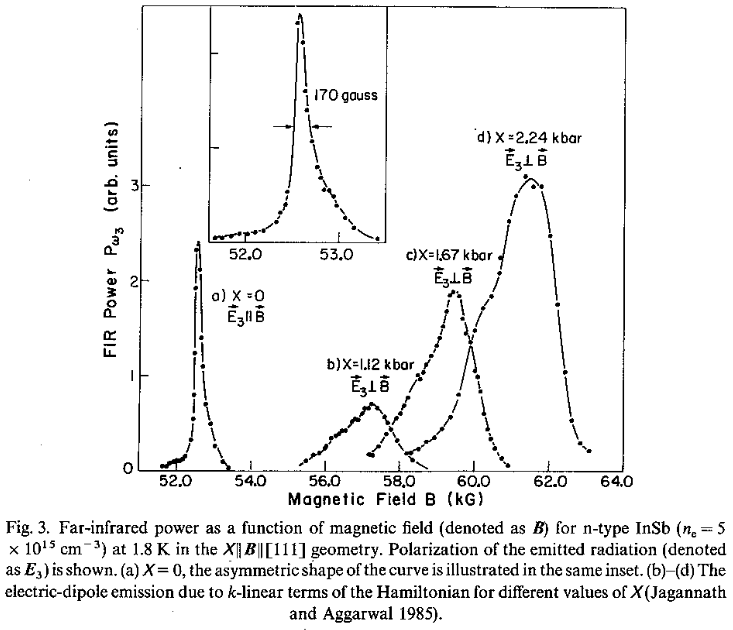} 
\caption{ 
} 
\label{fig3}
%\vspace{-4mm}
\end{figure}

 \begin{figure}[hbt!]
\includegraphics[width=0.95\columnwidth]{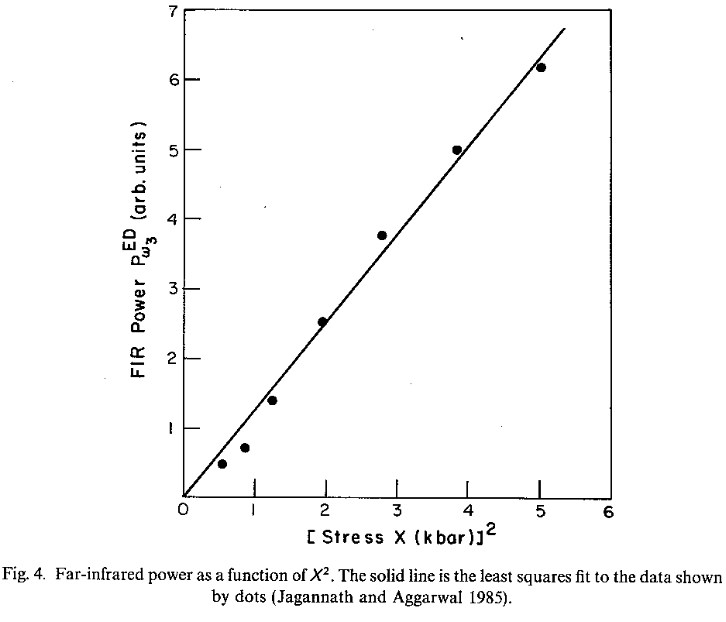} 
\caption{ 
} 
\label{fig4}
%\vspace{-4mm}
\end{figure}

 \begin{figure}[hbt!]
\includegraphics[width=0.95\columnwidth]{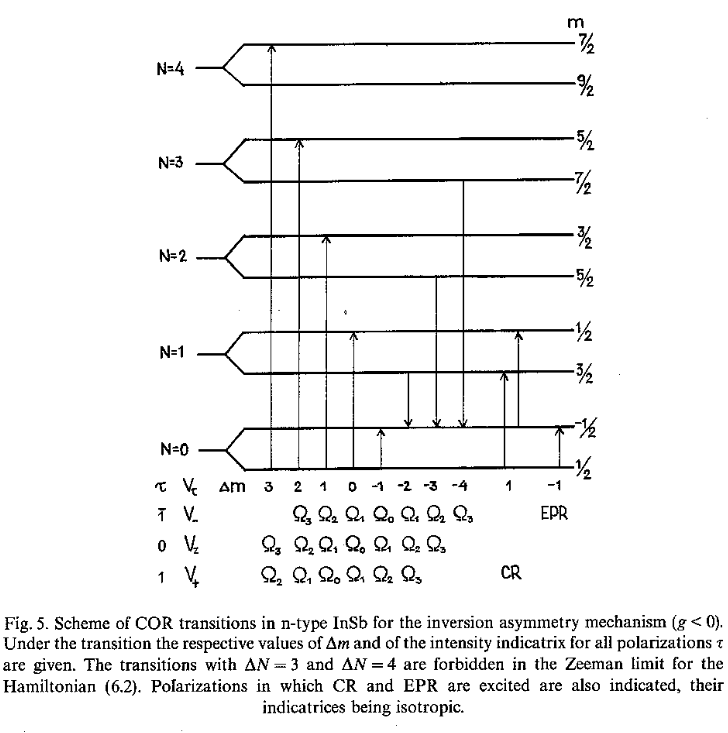} 
\caption{
} 
\label{fig5}
%\vspace{-4mm}
\end{figure}

 \begin{figure}[hbt!]
\includegraphics[width=0.95\columnwidth]{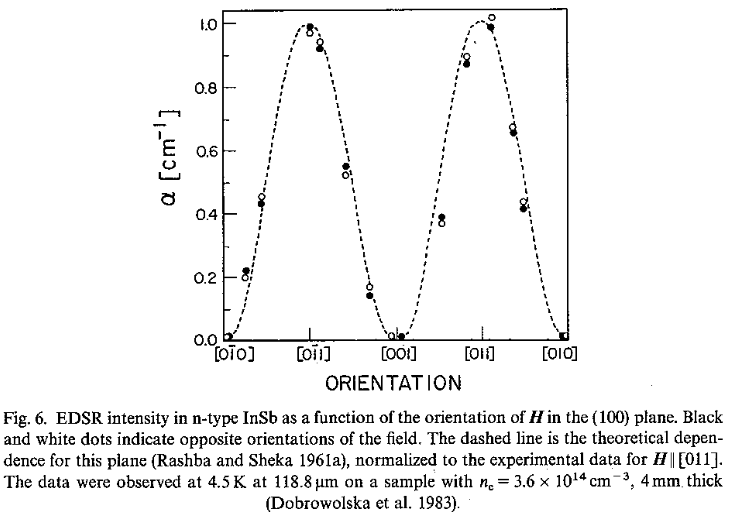} 
\caption{
} 
\label{fig6}
%\vspace{-4mm}
\end{figure}

 \begin{figure}[hbt!]
\includegraphics[width=0.95\columnwidth]{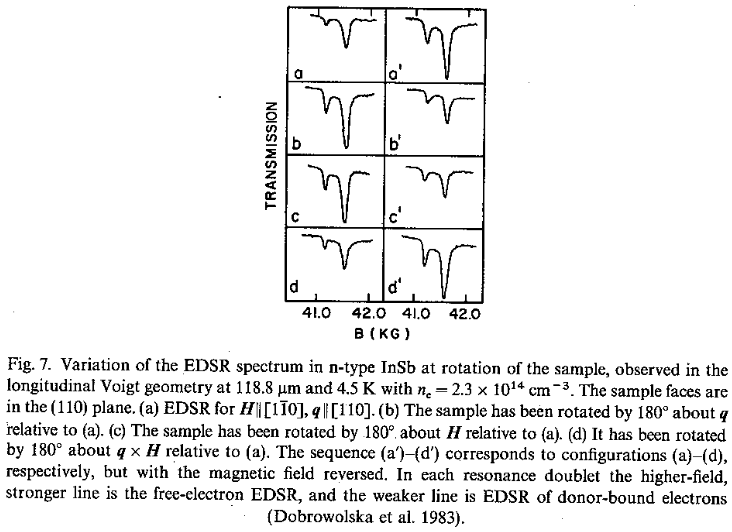} 
\caption{
} 
\label{fig7}
%\vspace{-4mm}
\end{figure}

 \begin{figure}[hbt!]
\includegraphics[width=0.95\columnwidth]{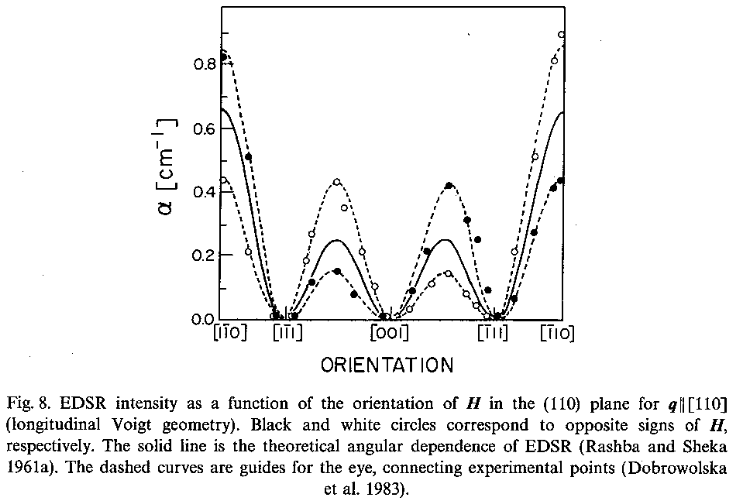} 
\caption{
} 
\label{fig8}
%\vspace{-4mm}
\end{figure}

 \begin{figure}[hbt!]
\includegraphics[width=0.95\columnwidth]{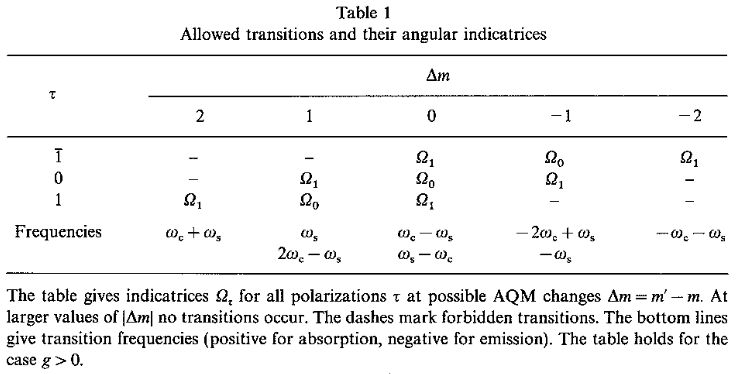} 
\caption{
} 
\label{fig9}
%\vspace{-4mm}
\end{figure}

\medskip

\newpage
{\it References:}

Abrikosov. A.A., and L.A. Fal'kovskii, 1962, Zh. Eksp. Tcor. Fiz. 43, 1089 (Sov. Phys-JETP 16, 769).

Appold, G., H. Pascher, R. Ehen, U. Steigenbergen and M. von Ortenberg, 1978. Phys. Status Solidi В 86, 557.

Aronov, A.G.. G.R. Pikus and A.N. Titkov. 1983, Zh. Eksp. Teor. Fiz. 84, 1170 (Sov. Phys-JETP 57,  680).

Babich, V.M., N.P. Baran, A.A. Bugaj, A.A. Konchitz and B.D. Shanina, 1988, Zh. Eksp. Teor. Fiz.
94. 319 (Sov. Phys.-J ETP 67, 1697).

Balkanski, M, and J. Cloizeaux, 1960, J. Phys, et Radium 21, 825.

Bangert, E., 1981, L. N. Phys. (Springer-Verlag, Berlin) 152, 216.

Bell, R.L.. 1962, Phys. Rev. Lett. 9, 52.

Bell, R.L., and K.T. Rogers, 1966. Phys. Rev. 152, 746.

Bir. G.L., and G.E. Pikus, 1959, Fiz. Tverd. Tela 1, 1642.

Bir, G.L., and G.E. Pikus, 1961, Fiz. Tverd. Tela 3, 3050 (1962, Sov. Phys.-Solid State 3, 2221).

Bir, G.L., and G.E. Pikus, 1972, Symmetry and Strain-Induced Effects in Semiconductors (Nauka, Moscow, English translated, 1974. Halsted Press, New York).

Bir, G.L., E.I. Butikov and G.E. Pikus, 1963, J. Phys. Chcm. Solids 24, 1475.

Bloembergen, N.. 1961, Sei. 133, 1363.

Blount, E.I., 1962. Solid State Phys.. eds F. Seitz and D. Turnbull (Academic Press. New York) 13, 305.

Bodnar, J., 1978, in: Phys. Narrow-Gap Scmicond., eds J. Rauluszkiewicz, M. Gorska and E. Kaczmarek (Elsevier, Amsterdam; PWN. Warsaw) p. 311.

Boiko, I. I., 1962, Fiz. Tverd. Tela 4. 2128 (Sov. Phys.-Solid State 4, 1558).

Boiko, I. I., 1964. Ukr. Fiz. Zh. 9, 1256.

Bowers, R., and Y. Yafet, 1959, Phys. Rev. 115, 1165.

Braun, M., and U. Rössler, 1985, J. Phys. C: Solid State Phys. 18, 3365.

Brueck, S. R. J., and A. Mooradian. 1973, Opt. Commun. 8, 263.

Burgiel, J. C., and L. C. Hebei, 1965, Phys. Rev. A 140, 925.

Bychkov, Yu.A., and E.I. Rashba. 1984, Pis'ma Zh. Eksp. Teor. Fiz. 39, 66 (Sov. Phys.-JETP Lett. 39, 78).

Bychkov, Yu. A., and E. I. Rashba, 1985, in: Proc. 17th Int. Coni. Phys. Scmicond., San Francisco, 1984, eds J.D. Chadi and W.A. Harrison (Springer-Verlag, New York) p. 321.

Cardona, M., N. E. Christensen and G. Fasol, 1986a, Phys. Rev. Lett. 56, 2831.

Cardona, M., N. E. Christensen, M. Dobrowolska, J. K. Furdyna and S. Rodriguez, 1986b, Solid State Commun. 60, 17.

Cardona, M., N. E. Christensen and G. Fasol, 1987, in: Proc. 18th Int. Conf. Phys. Scmicond..
Stockholm, 1986, p. 1133.

Casella, R.C., 1960, Phys. Rev. Lett. 5, 371.

Chen, Y.-E, M. Dobrowolska and J.K. Furdyna, 1985a, Phys. Rev. В 31, 7989.

Chen, Y.-E, M. Dobrowolska, J.K. Furdyna and S. Rodriguez, 1985b, Phys. Rev. В 32, 890.

Cohen. M.H., and EЛ. Blount, 1960, Philos. Mag. 5, 115.

Curie, P, 1894, J. Phys. (3rd series) 3, 393.

Darr, A., J. P. Kotthaus, T. Ando, 1976, in: Proc. 13th Int. Conf. Phys. Semicond., Rome, 1976, ed. F. G. Fumi (Marvcs, Rome) p. 774.

Das, B., U. C. Miller, S. Datta. R. Reifenberger. W. R. Hong, P. K. Bhattacharva. J. Singh and M. Jaffe, 1989, Phys. Rev. B 39, 1411.

Dickey, D. H., and D.M. Larsen, 1968, Phys. Rev. Lett. 20, 65.

Died, T., 1983, J. Magn. Magn. Mater. 38, 34.

Dobers, M., K. von Klitzing and G. Weimann, 1988, Phys. Rev. B 38, 5453.

Dobrowolska, M., H. D. Drew, J. K. Furdyna, T. Ichiguchi, A. Witowski and P. A. Wolff. 1982, Phys. Rev. Lett. 49, 845.

Dobrowolska, M., Y. Chen, J.K. Furdyna and S. Rodriguez, 1983, Phys. Rev. Lett. 51, 134. 

Dobrowolska, M., A. Witowski, J.K. Furdyna, T. Ichiguchi, H.D. Drew and P.A. Wolff, 1984, Phys. Rev. B 29, 6652.

Dorozhkin, S. I., and E.B. Ol'shanetskii, 1987, Pis'ma Zh. Eksp. Teor. Fiz. 46, 399 (Sov. Phys.-JETP Lett. 46, 000).

Edelstein, V.M., 1983, Solid State Commun. 45, 515.

Elliott. R.J., 1954, Phys. Rev. 96, 280.

Erhardt, W., W. Staghuhn, P. Byszewski, M. von Ortenberg, G. Landwehr, G. Weimann, L .van Bockstal, P. Janssen, F. Herlach and J. Witters, 1986. Surf. Sci. 170, 581.

Fantner, E. J., H. Pascher, G. Bauer, R. Danzer and A. Lopez-Otero, 1980. J. Phys. Soc. Jpn. 49 Suppl. A, 741.

Gatos, H. C., and M. C. Levine, 1960, J. Phys. Chem. Solids 14, 169.

Gershenzon, E. M., N. M. Pevin and M. S. Fogel'son, 1970, Pis'ma Zh. Eksp. Teor. Fiz. 12,201 (Sov.
Phys.-JETP Lett. 12, 139).

Gershenzon, E. M., N. M. Pevin, I. T. Semenov and M. S. Fogel'son, 1976, Fiz. Tekh. Polupr. 10, 175 (Sov. Phys.-Semicond. 10, 104).

Glinchuk, M. D., V. G. Grachev, M. F. Deigen, A. B. Roitisin and L. A. Suslin. 1981, Electric Effects in Radiospectroscopy (in Russian), Nauka, Moscow.

Golin, S.. 1968, Phys. Rev. 166, 643.

Golubev, V. G., and V. I. Ivanov-Omskii, 1977. Pis'ma Zh. Tekh. Fiz. 3. 1212 (Sov. Tech. Phys. Lett. 3, 501).

Gopalan, S., J. K. Furdyna and S. Rodriguez, 1985. Phys. Rev. В 32, 903.

Gopalan, S., S. Rodriguez, J. Myeielski, A. Witowski, M. Grynberg and A. Wittlin, 1986, Phys. Rev. В 34, 5466.

Grisar, R.. H. Wachernig, G. Bauer, S. Hayashi, E. Amzallag, J. Wlasak and W. Zawadzki, 1976, in: Proc. I3th Int. Conf. Phys. Semicond.. Rome, 1976, ed. KG. Fumi (Marvcs, Rome) p. 1265. 

Gurgenishvili. G. E.. 1963, Fiz. Tverd. Tela 5. 2070 (Sov. Phys.-Solid State 5, 1510).

Hensel, J. C. 1968, Phys. Rev. Lett. 21, 983.

Hermann, C, and G Weisbuch, 1977. Phys. Rev. B 15, 823.

Ivchenko, E. L., and A. V. Sel'kin, 1979, Zh. Eksp. Teor. Fiz. 76, 1837 (Sov. Phys.-JETP 49, 933). 

Jagannath, C., and R.L. Aggarwal, 1985, Phys. Rev. B 32, 2243.

Johnson, E. J., and D.M. Larsen. 1966, Phys. Rev. Lett. 16, 655.

Kaenian, P., and W. Zawadzki, 1976. Solid State Commun. 18, 945,

Kalashnikov, V. H., 1974, Teor. Matern. Fiz. 18, 108.

Kane, E. O., 1957, J. Phys. Chem. Solids 1, 249.

Kohn. W., 1957, in: Solid State Phys., eds F. Seitz and I). Turnbull (Academie Press, New York) 5, 257.

Koshelev. A. E., V. Ya. Kravchenko and D. E. Khmel'nitskii, 1988, Fiz. Tverd. Tela 30, 433. 

Kriechbaum. M.. R. Meisels, F. Kuchar and E. Fantner. 1983. in: Proc. 16th Int. Conf. Phys.
Semicond., Montpellier, 1982, cd. M. Averous (North-Holland, Amsterdam) p. 444.

Kuchar, F., R. Meisels, R.A. Stradling and S.P. Najda, 1984, Solid State Commun. 52, 487.

Kveder, V. V., Yu. A. Osipyan and А. I. Shalynin, 1984, Pis'ma Zh. Eksp. Teor. Fiz. 40, 10 (Sov.
Phys.-JETP Lett. 40, 729).

Kveder, V. V., V. Ya. Kravchenko, T. R. Mchedlidze, Yu. A. Osip'yan, D. E. Khmel'nitskii and A. I. Shalynin, 1986, Pis'ma Zh. Eksp. Teor. Fiz. 43, 202 (Sov. Phys.-JETP Lett. 43, 255).

Kveder, V. V., A. E. Koshelev, T. R. Mchedlidze, Yu. A. Osip'yan and A.I. Shalynin, 1989, Zh. Eksp.
Teor. Fiz. 95, 183 (Sov. Phys.-JETP 68, 104).

La Rocca, G.C., N. Kim and S. Rodriguez, 1988a, Phys. Rev. B 38, 7595.

La Rocca, G.C., N. Kim and S. Rodriguez, 1988b, Solid State Commun. 67, 693.

Landau, L. D., and E. M. Lifshitz, 1974, Quantum Mechanics. Nauka (English translated, 3rd edition, Pergamon Press, 1977).

Lax, B., J.G. Mavroides, H.J. Zeiger and R.J. Keyes. 1961, Phys. Rev. 122, 31.

Leibier, L. 1978, Phys. Status Solidi В 85, 611.

Lin-Chung, P.J., and B.W. Henvis, 1975, Phys. Rev. В 12, 630.

Littler, C.L., D. G. Seiler. R. Kaplan and R. J. Wagner, 1983, Phys. Rev. B 27, 7473.

Lommer, G., F. Malcher and U. Rossler. 1985, Phys. Rev. B 32, 6965.

Ludwig, G.W., and F.S. Ham, 1962, Phys. Rcv. Lett. 8, 210.

Luo, J., H. Munekata, F.F. Fang and P.J. Stiles, 1988, Phys. Rev. B 38, 10142.

Luttinger, J.M., 1956, Phys. Rev. 102, 1030.

Luttinger, J.M., and W. Kohn, 1955, Phys. Rev. 97, 869.

Malcher, F, G. Lommer and U. Rossler, 1986, Superlattices and Microstructures 2, 267.

McClure, J.W., and K.H. Choi, 1977. Solid State Commun. 21, 1015.

McCombe, B.D., 1969. Phys. Rev. 181, 1206.

McCombe, B.D., and R. Kaplan, 1968, Phys. Rev. Lett. 21, 756.

McCombe, B.D., and R.J. Wagner, 1971, Phys. Rev. В 4, 1285.

McCombe, B.D., S.G. Bishop and R. Kaplan, 1967, Phys. Rev. Lett. 18, 748.

McCombe, B.D., R.J. Wagner and G.A. Prinz, 1970a, Solid State Commun. 8, 1687.

McCombe, B.D., R.J. Wagner and G.A. Prinz, 1970b, Phys. Rev. Lett. 25, 87.

McCombe, B.D., R.J. Wagner and J.S. Lannin, 1974, in: Proe. 12th Int. Conf. Phys. Semicond..
Stuttgart, 1974 (B.G. Teubner, Stuttgart) p. 1176.

Mel'nikov, V.I., and E.I. Rashba, 1971, Zh. Eksp. Teor. Fiz. 61, 2530 (1972. Sov. Phys.-JETP 34, 1353).

Merkt, U., M. Horst, T. Evelbauer and J.P. Kotthaus, 1986, Phys. Rev. 34, 7234.

Mims, W.B., 1976, The Linear Electric Field Effect in Paramagnetic Resonance (Clarendon Press, Oxford).

Nguyen, V.T., and T.J. Bridges, 1972. Phys. Rev. Lett. 29, 359.

O'Dell, T.H., 1970, The Electrodynamics of Magnetoelectric Media (North-Holland, Amsterdam). 

Ogg, N.R., 1966, Proc. Phys. Soc. 89, 431.

Ohta, K., 1969, Phys. Rev. 184, 721.

Pascher, H., 1981, L. N. Phys. (Springer-Verlag, Berlin) 152, 202.

Pastor К. M. Jaczynski and J.K. Furdyna, 1981. Phys. Rev. В 24, 7313.

Pekar, S.I., and E. I. Rashba, 1964, Zh. Eksp. Teor. Fiz. 47, 1927 (1965, Sov. Phys.-JETP 20, 1295). 

Pevtsov, A.B., and A.V. SelTin. 1983, Fiz. Tverd. Tela 25, 157 (Sov. Phys.-Solid State 25, 85). 

Pidgeon, C.R., and R.N. Brown, 1966, Phys. Rev. 146, 575.

Pidgeon, C.R., and S.H. Groves, 1969, Phys. Rev. 186, 824.

Ranvaud, R. H.-R. Trebin, U. Rossler and F.H. Poliak, 1979, Phys. Rev. В 20, 701.

Rashba. E. I., I960. Fiz. Tv. Tela 2, 1224 (Sov. Phys.-Solid State 2, 1109).

Rashba, E.I., 1961. in: Proc. Int. Conf. Phys. Semicond., Prague. I960 (Publ. House of the Czechosl. Acad. Sei., Prague) p. 45.

Rashba, E.I.. 1964a, Usp. Fiz. Nauk 84, 557 (1965, Sov. Phys.-Usp. 7, 823).

Rashba. E.I., 1964b. Fiz. Tverd. Tela 6, 3178 (1965, Sov. Phys.-Solid State 6, 2538).

Rashba. E.I., 1979, J. Magn. Magn. Mater. 11, 63.

Rashba, E.I., and VI. Sheka, 1959, Fiz. Tverd. Tela, selected papers. Vol. 2, 162.

Rashba, E.I.. and VI. Sheka, 1961a, Fiz. Tverd. Tela 3, 1735 (Sov. Phys.-Solid State 3. 1257). 

Rashba, E.I. and VI. Sheka, 1961b, Fiz. Tverd. Tela 3, 1863 (Sov. Phys.-Solid State 3, 1357). 

Rashba, E.I. and V.I, Sheka, 1961c. Fiz. Tverd. Tela 3, 2369 (Sov. Phys.-Solid State 3, 1718). 

Rashba, E.I., and VI. Sheka, 1964a. Fiz. Tverd. Tela 6, 141 (Sov. Phys.-Solid State 6, 114).

Rashba, E.I., and VI. Sheka, 1964b, Fiz. Tverd. Tela 6.576 (Sov. Phys.-Solid State 6, 451).

Roitsin, A.B.. 1971, Usp. Fiz. Nauk, 105. 677 (1972, Sov. Phys.-Usp. 14, 766).

Romestain. R- S. Geschwind and G.E. Devlin, 1977, Phys. Rev. Lett. 39, 1583.

Rossler, U., 1984, Solid State Commun. 49, 943.

Roth, L.M., B. Lax and S. Zwerdling, 1959, Phys. Rev. 114, 90.

Rubo, Yu. G.. L.S. Khasan, V.I. Sheka and A.S. loselcvich. 1988, Pis'ma Zh. Eksp. Teor. Fiz. 48,30 (Sov. Phys.-JETP Lett. 48, 30).

Rubo, Yu. G., L. S. Khasan, V. I. Sheka and E. V. Mozdor, 1989, Zh. Eksp. Teor. Fiz. 95. 1180 (Sov.
Phys.-JETP 68, 1087).

Schaber, FL and R.E. Doezema, 1979a, Solid State Commun. 31, 197.

Schaber, H., and R.E. Doezema, 1979b, Phys. Rev. В 20, 5257.

Seiler, D. G., W. M. Becker and L.M. Roth, 1970, Phys. Rev. В 1, 764.

Sheka. VI.. 1964, Fiz. Tverd. Tela 6, 3099 (Sov. Phys.-Solid State 10, 2470).

Sheka, VI., and L.S. Khazan, 1985, Pis'ma Zh. Eksp. Teor. Fiz. 41, 61 (Sov. Phys.-JETP Lett. 41, 72).

Sheka, V.I., and I.G. Zaslavskaya, 1969, Ukr. Fiz. Zh. 14, 1825.

Singh. M., and P.R. Wallace. 1983, Physica В 117 \& 118, 441.

Slusher, R.E., C.K.N. Patel and P.A. Fleury. 1967. Phys. Rev, Lett. 18, 77.

Smith, G.E., J.K. Galt and F.R. Merritt, I960. Phys. Rev. Lett. 4, 276.

Stein, D.. K. von Klitzing and G. Weimann, 1983. Phys. Rev. Lett. 51, 130.

Stepniewski, R., 1986, Solid State Commun. 58, 19.

Stepniewski, R., and M. Grynberg, 1985, Acta Phys. Pol. A 67, 373.

Stormer, H.L., 1988, Materials of the Soviet-American Seminar: Electronic Properties of two-dimensional systems (Moscow. 1988), unpublished.

Sugihara, K., 1975, J. Phys. Soc. Jpn. 38, 1061.

Thielemann, J., M. von Ortenberg, F.A.P. Blom and К. Strobel, 1981, L. N. Phys. (Springer-Verlag, Berlin) 152, 207.

Trebin, H.-R., U. Rössler and R. Ranvaud, 1979, Phys. Rev. В 20, 686.

Tuchendler, J.. M. Grynberg, Y. Couder, H. Thome and R. Le Toullec, 1973, Phys. Rev. В 8, 3884. 

Verdun, H R., and H.D. Drew, 1976, Phys. Rev. В 14, 1370.

Weiler, M.H., 1982, Solid State Commun. 44, 287.

Weiler, M.H., R.L. Aggarwal and B. Lax, 1978, Phys. Rev. В 17, 3269.

Wigner, E.P., 1959, Group Theory and its Application to the Quantum Mechanics of Atomic Spectra (Academic Press, New York).

Witowski, A., K. Pastor and J.K. Furdyna, 1982, Phys, Rev. В 26, 931.

Wlasak, J., 1986, J. Phys. C: Solid State Phys. 19, 4143.

Wolff, P. A., 1964, J. Phys. Chem. Solids 25, 1057.

Yafet, Y., 1963, in: Solid State Phys., eds F. Seitz and D. Turnbull (Academic Press, New York) 14,
1.

Yafet, Y.. 1966, Phys. Rev. 152, 858.

Zavoisky, E.K., 1945, J. Phys. USSR 9, 245.

Zawadzki, W., and J. Wlasak, 1976, .1. Phys. C: Solid State Phys. 9, L663.

Zawadzki, W., P. Pfeffer and H. Sigg, 1985, Solid State Commun. 53, 777.

\end{document}